\documentclass{kapproc} 

\usepackage{txfonts}
\usepackage{procps} 
\usepackage[dvips]{graphicx}

\upperandlowercase
\kluwerbib 

\begin{document}
\newcommand{\hi} {\rm H\,{\rm I}}
\newcommand{\hii} {\rm H\,{\rm II}}
\newcommand{\ha} {\rm H \alpha}
\newcommand{\kms} {\,{\rm km\,s}^{-1}}
\newcommand{\kpc} {\,{\rm kpc}}
\newcommand{\de}{^{\circ}}
\newcommand{\mo}{\,{M}_\odot}
\newcommand{\yr}{\,{\rm yr}}
\newcommand{\moyr}{\,{M_\odot\,\rm yr}^{-1}}
\newcommand{\mopc}{M_\odot\,{\rm pc^{-2}}}
\newcommand{\lo}{L_{\odot}}
\newcommand{\loB}{L_{\odot \rm, B}}
\newcommand{\gsim}{\lower.7ex\hbox{$\;\stackrel{\textstyle>}{\sim}\;$}}
\newcommand{\lsim}{\lower.7ex\hbox{$\;\stackrel{\textstyle<}{\sim}\;$}}
\newcommand{\ergs}{\,{\rm erg\,s}^{-1}}
\newcommand{\e}{{\rm e}}

\articletitle[]{The gaseous haloes of disc galaxies}

\author{Fraternali F.\altaffilmark{1}, Oosterloo T.\altaffilmark{2}, 
Binney J.J.\altaffilmark{1}, Sancisi R.\altaffilmark{3,4}}

\altaffiltext{1}{Theoretical Physics, University of Oxford (UK)}
\altaffiltext{2}{ASTRON, Dwingeloo (NL)}
\altaffiltext{3}{INAF-Osservatorio Astronomico, Bologna (I)}
\altaffiltext{4}{Kapteyn Astronomical Institute, Groningen (NL)}

\begin{abstract}

The study of gas outside the plane of disc galaxies is crucial
to understanding the circulation of material within a galaxy and
between galaxies and the intergalactic environment.
We present new $\hi$ observations of the edge-on galaxy NGC\,891,
which show an extended halo component lagging behind the disc in rotation.
We compare these results for NGC\,891 with other detections of gaseous
haloes.
Finally, we present a dynamical model for the formation of extra-planar
gas. 

\end{abstract}

\section{Introduction}

Gas plays a vital role in the evolution of disc galaxies.
The collapse of gas produces star formation; stars enrich gas, 
expel it via winds and supernova explosions and create a circulation 
called the galactic fountain (e.g.\ \cite{sha76}).
Outflows of gas from galactic discs with velocities of the order of
$100 \kms$ are indeed observed both in the neutral
(\cite{kam91}) and the ionised phase (\cite{fra04}).
On the other hand, disc galaxies also seem to acquire a substantial 
amount of gas from their surroundings. 
Accretion of unpolluted material from the
Intergalactic Medium (IGM) is predicted by chemical evolution models
of the Milky Way (e.g.\ \cite{roc96}) and
the low metallicities of some of the High Velocity Clouds (HVCs) 
(e.g.\ \cite{tri03})
suggest that cold material is indeed accreted by spiral galaxies like
our own (\cite{oor70}).

In recent years, a new and important fact has emerged. 
Observations at various wavelengths
have shown the presence of a considerable amount of gas in the 
haloes of disc galaxies (extra-planar or halo gas). 
Deep $\ha$ observations have revealed the presence of extended layers  
of diffuse ionised gas (DIG) around edge-on spiral galaxies
(\cite{hoo99}).
X-ray observations
have shown the presence of hot plasma at distances of tens of kpc from 
the plane of galactic discs (e.g.\ \cite{str04}).
$\hi$ observations have revealed extended
thick layers surrounding the galactic discs (\cite{swa97,fra01,mat03}).
The origin and the nature of this halo gas are still unknown.
However, it is a unique probe to study the exchange of gas
within a galaxy and between galaxies and the IGM.

\section{NGC\,891}

NGC\,891 is one of the best known and studied nearby edge-on 
galaxies.
It is at the distance of 9.5 Mpc, classified as 
a Sb/SBb, and it is often referred to as 
very similar to the Milky Way (e.g.\ \cite{vdk81}).

\begin{figure}[ht]
\begin{center}
\includegraphics[width=110mm]{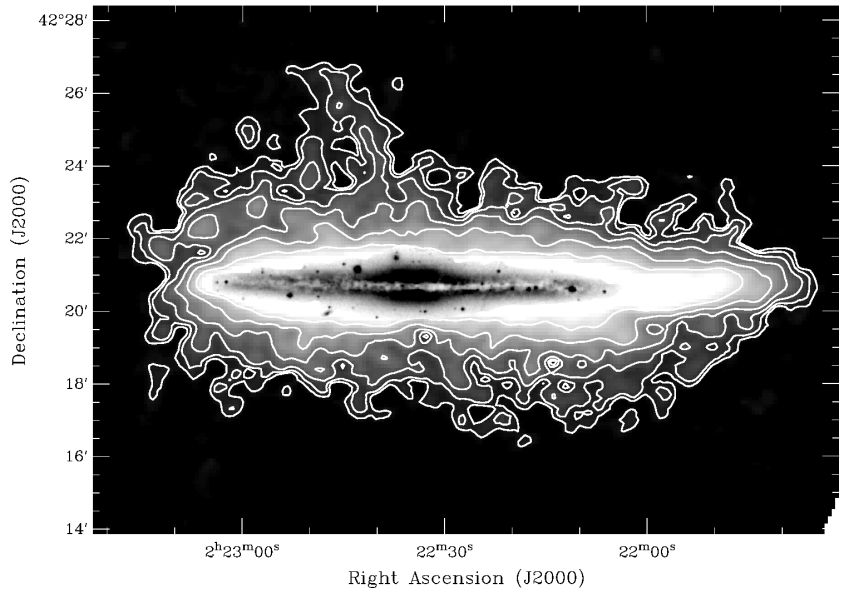}
\caption{
\protect
Optical DSS image (grey-scale) and total $\hi$ map 
(contours$+$negative grey-scale) of NGC\,891, 
the latter obtained from the new WSRT observations
(\cite{oos05}).
$\hi$ contours are: 1.7, 4.5, 9, 18.5, 37, 74, 148, 296.5, 593 $\times$
10$^{19}$ atoms cm$^{-2}$. The beam size is $28''$; 1$'=$2.8 kpc.
}
\end{center}
\end{figure}

NGC\,891 has been the subject of numerous
studies at different wavelengths that have led to the detection of
various halo components:
an extended radio halo (\cite{all78}),
a thick layer of diffuse ionised gas (DIG) (e.g.\ \cite{det90}),
and diffuse extra-planar hot gas (\cite{bre94}).
Also ``cold'' ISM components have been detected in the 
halo such as $\hi$ (\cite{swa97}), dust (\cite{how99}) and
CO (\cite{gar92}).

New $\hi$ observations of NGC\,891 have been obtained with the
Westerbork Synthesis Radio Telescope (WSRT) with a total integration
time of about 200 hrs (\cite{oos05}).
Fig.\ 1 shows the new total $\hi$ map of NGC\,891 in contours overlaid
on a grey-scale DSS optical image.
This $\hi$ halo is remarkably extended with a spur of gas
reaching a distance of 15 kpc from the plane.
The $\hi$ disc of NGC\,891 is lop-sided, more extended on the S-W side of 
the galaxy (right in Fig.\ 1) and possibly $truncated$ on the N-E
(left) side (see also \cite{san79}; \cite{swa97}).

The halo gas in NGC\,891 rotates more slowly than the gas in the plane
(\cite{swa97}).
This is a common property of gaseous haloes (e.g.\ 
\cite{fra01,hea05a}).
Our new data allow us to study the kinematics of 
the halo gas in detail and derive
rotation curves at different heights from the plane
(\cite{fra05a}; Fraternali, in preparation).
The measured gradient in rotation velocity is $\Delta_{\rm v_{\rm rot}} 
\sim -15 \kms {\rm kpc}^{-1}$ for 1.3$<$z$<$5.2 kpc. 
This value agrees with the velocity 
gradient measured in the ionised gas (\cite{hea05b}).

\section{Halo gas in other disc galaxies}

How common are gaseous haloes? 
Until now, neutral extra-planar gas has been studied in 6 disc galaxies.
The $\hi$ data for these galaxies are among the deepest ever obtained
and this may indicate that halo gas is a common feature
that has escaped detection because of a lack of sensitivity.

\begin{table}[ht]
\caption[Observations of halo gas in disc galaxies
]
{Observations of halo gas in disc galaxies}
\begin{tabular*}{\textwidth}{@{\extracolsep{\fill}}lcccccc}
\sphline
Galaxy & Type & Distance & M$_{\rm halo}$ $\hi$ & M$_{\rm halo}$/M$_{\rm total}$ & L(FIR) & Ref.\ \cr
 & & (Mpc) & ($10^8~ \mo$) & (\%) & ($\lo$) \cr
\sphline
NGC\,253 & Sc & 3.9 & 0.8 & 3 & 2.63$\times 10^{10}$&1\cr
NGC\,891 & SBb & 9.5 & 6 & 15 & 1.25$\times 10^{10}$&2\cr
NGC\,6946 & Scd & 6.0 & $>$3.5 & $>$5.8 & 9.79$\times 10^9$ &3\cr
NGC\,4559 & Scd & 9.7 & 5.9 & 11.5 & 1.91$\times 10^9$&4\cr
NGC\,2403 & Scd & 3.2 & 3 & 11.0 & 1.13$\times 10^9$ &5\cr
UGC\,7321 & Sd & 10.0 & $\gsim 0.1$ & $\gsim$1 & 7$\times 10^7$ &6\cr
\sphline
\end{tabular*}
\begin{tablenotes}
$^1$ \cite{boo05a};
$^2$ \cite{swa97};
$^3$ \cite{boo05b};
$^4$ \cite{barb05};
$^5$ \cite{fra02};
$^6$ \cite{mat03}.
\end{tablenotes}
\end{table}
\inxx{captions,table}

Table 1 summarizes the 6 halo gas detections 
sorted by their FIR luminosity (which is a measure of the star
formation rate, SFR). 
A relation between the amount of halo gas and the SFR
may be expected if the halo gas has a galactic fountain
origin. 
It is possible that a trend is indeed present
for low luminosity galaxies. 
Note that the mass of halo gas estimated for NGC\,6946 is a
lower limit due to the low inclination ($\sim30 \de$) 
of the galaxy that does not allow
an efficient separation of the halo component (\cite{boo05b}).
However, the starburst galaxy NGC\,253 is totally anomalous, having
very little extra-planar $\hi$ compared to the other galaxies.
This may be an indication 
that in the most actively star forming galaxies most of
the halo gas is ionised.

In addition to these galaxies, there are indications of the presence of 
halo gas in several others (e.g.\ NGC 5055, \cite{bat05}; UGC\,1281, 
Kamphuis et al., this conference; UGC\,12632, Oosterloo, private 
communication) and extra-planar features have been observed in 
NGC\,2613 (\cite{cha01}).
Moreover, it is very likely that this halo gas is analogous to the 
Intermediate
and High Velocity Clouds (IVCs/HVCs) of the Milky Way (\cite{wak97}) and 
of external galaxies 
(M31 and M33, \cite{wes05}; M83 and M51, 
\cite{mil05}).

\section{A dynamical model for extra-planar gas}

What is the origin of these gaseous haloes?
Some authors have tried to explain them using
ballistic galactic fountain models and compared them
with $\ha$ observations of ionised gas (\cite{col02, hea05a}).
They found a general disagreement between the kinematics and distribution
of the halo gas predicted by the models and those shown by the data. 
Other authors have tried to model the halo gas as a
stationary medium in hydrostatic equilibrium (\cite{ben02, barn05}).
This approach leads to solutions that can reproduce
the observed gradient in the rotation velocity (\cite{barn05}),
however the temperatures of the extra-planar (hydrostatic) gas are
of the order of 10$^5$ K and it is unclear how this medium can
be related to cold neutral gas.

We have developed a model for the formation of the neutral 
extra-planar gas that takes into account internal and external 
processes. 
At the current stage (\cite{fra05b}) the model tries to reproduce
the halo gas
as a continuous flow of collision-less particles from the disc into the halo 
region.
The particles are stopped at the passage through the plane.
The output of the models are particle-cubes that can be directly
compared to the \hi\ data-cubes.
The idea behind this approach is to start from a basic model, study 
its failures and improve it progressively to match the data more and more
closely.

\begin{figure}[ht]
\begin{center}
\includegraphics[width=110mm]{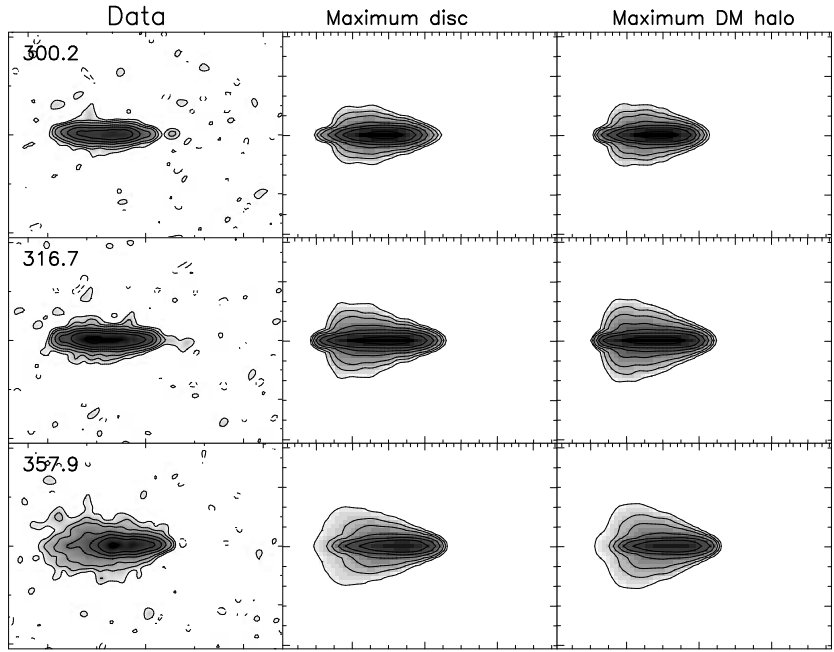}
\caption{
\protect
    Comparison between three observed channel maps of NGC\,891 (\cite{oos05})
    and those produced with the two dynamical models.
    The first column shows the data, heliocentric radial velocities
    are reported in the upper left corner. 
}
\end{center}
\end{figure}

Figure 2 shows three representative channel maps of NGC\,891 (left column)
compared with those produced by two dynamical models.
The two models are for maximum and minimum disc (maximum DM halo) 
potentials.
The channel maps shown here have velocities far from systemic
i.e.\ they show the gas that rotates the fastest.
The data show that this gas is mostly located in the disc
(the channel maps are thin) whilst
the models predict much thicker channel maps, i.e.\ fast
rotating extra-planar gas.
This is a failure of the fountain model
and this problem cannot be removed by allowing the particles to cross
the disc or by considering that they are ionised as they leave the plane
and become visible at 21$\,$cm at some point on their orbits 
(phase change).
There is an intrinsic and unavoidable 
need for low angular momentum material, which suggests that one
has to take into account interactions between the fountain flow and a
hot gaseous halo and/or accretion material 
(Fraternali \& Binney, in preparation).

\section{Conclusions}

The new $\hi$ observations of NGC\,891 show the presence of extended 
neutral gas halo reaching up the distance of 15 kpc from the plane.
This halo gas rotates more slowly than the gas in the plane (with a 
gradient of $-$15 $\kms {\rm kpc}^{-1}$). 
These kinematic properties cannot be explained by an isolated galactic 
fountain; they require that
interactions with a hot halo and/or gas accretion 
from the IGM must play an important role.

\bibliographystyle{kapalike}
\chapbblname{chapbib}
\chapbibliography{logic}

\end{document}